\documentstyle[aps,preprint]{revtex}
\newcommand{\be}{\begin{equation}}
\newcommand{\ee}{\end{equation}}
\newcommand{\ba}{\begin{eqnarray}}
\newcommand{\ea}{\end{eqnarray}}
\newcommand{\nn}{\nonumber}
\newcommand{\gm}{\gamma}

\newcommand{\plb}{Phys. Lett. B$~$}

\newcommand{\npb}{Nucl. Phys. B}
\begin{document}


\thispagestyle{empty}

\title{
\hbox to\hsize{\large Submitted to Phys.~Lett.~B \hfil E-Print
hep-ph/980000}
\vskip1.55cm
Axion Emissivity from the Conversion 
of a Neutron Star into a Strange Star}

\author{
In-Saeng Suh
\footnote{E-mail: InSaeng.Suh.6@nd.edu,$~~$life@hepth.hanyang.ac.kr}
, Chul H. Lee 
\footnote{E-mail: chlee@hepth.hanyang.ac.kr}
}
 
\vspace{0.8 cm}

\address{
The Research Institute for Natural Sciences and
Department of Physics, \\
Hanyang University, 133-791, Seoul, Korea}

\vspace{0.3 cm}

\date{\today}
\maketitle

\begin{abstract} 
The conversion of neutron matter into strange matter in a neutron star
occurs through the non-leptonic weak-interaction process.
We study the energy loss of the neutron star by the emission of axions
in that process. Owing to that process, the neutron star will liberate
the energy which can in no way be negligible as an axion burst.
\\
\\
\end{abstract}



\newpage

\pagenumbering{arabic}\setcounter{page}{1}

\section{Introduction}

The possibility of the existence of stable quark matter in the early universe
or inside a neutron star  or in relativistic heavy-ion
collision experiments has been studied in many works \cite{sqm}\@.
In this work, we consider the conversion of a neutron star into a strange star
and its energy loss during the conversion process.
In order to form strange matter in the interior of a neutron star, neutron 
matter should be converted into strange matter. Several conversion mechanisms 
have been discussed by Alcock et al.\ \cite{alcock}\@.
When a neutron and a stable strangelet (strange quark matter droplet) meet, 
the neutron is readily absorbed, 
while a proton can coexist with a strangelet due to the Coulomb 
barrier between them \cite{alcock}\@.
Therefore, a stable strangelet as a seed for the process will trigger
a conversion in such a way that it grows by absorbing neutrons and, eventually,
convert most of the neutron star into a strange star \cite{olinto,daietal}\@.
Then the conversion would liberate about $10^{52}$ ergs in binding energy 
\cite{olinto}\@.

A variety of mechanisms have been suggested for seeding the interior of a
neutron star with stable strangelets so far \cite{alcock,lugones}\@.
They can be divided into two main categories \cite{olinto};
(1) the primary mechanisms in which the seed is formed inside the neutron star,
(2) the secondary mechanisms in which the seed comes from the interstellar 
medium. In a viewpoint of the primary mechanisms \cite{alcock},
the central high densities and pressures in a neutron star may be sufficient 
for a phase transition from neutron matter to two-flavor quark matter to occur. 
Subsequently two-flavor quark matter can then easily decay into the lower 
energy strange matter through weak interactions.

If we assume that strangelet has formed inside a neutron star,
subsequent conversion process of the rest of the star is described
as follows \cite{olinto,olesen}\@.
The volume over which strange matter equilibrates was shown to be much 
smaller than that of the total strange matter region \cite{lugones,horvath}, 
so that the problem can be treated hydrodynamically in one-dimensional geometry.
As the conversion front sweeps into neutron matter,
the small region behind the conversion front 
has an excess of down quarks relative to strange quarks 
due to the flux of neutrons ($udd$) at the conversion front.
The excess down quarks will convert into strange quarks via the non-leptonic 
weak process \cite{madsen}, 
$d \; + \; u \; \rightarrow \; s \; + \; u$, as 
long as $\mu_d > \mu_s (n_d > n_s)$\@.
By this process a $d$ quark can change itself to an $s$ quark until
the Fermi energies of all the flavors become equal and the energy per baryon
drops in comparison to the ordinary two-component nuclear matter.
Other leptonic decay processes are suppressed considerably \cite{madsen}\@.
The conversion will liberate $\sim 10^{52}$ ergs of energy 
(assuming that $\sim$ 10 $MeV$ is liberated per neutron converted)\@.
This energy will be radiated as neutrinos, photons, $e^{+}e^{-}$ pairs, and
axions, etc. In this stage neutrinos and axions can escape the star.
Many authors \cite{daietal,aggs} have considered the escape of neutrinos.
In this work we calculate the energy loss due to the emission 
of axions in the process of such a conversion of a neutron star 
into a strange star.

At the chemical equilibrium between the quarks and the electrons, 
non- or semi-leptonic weak interaction is not important in neutrino and 
axion emission in quark matter
because the weak interactions coupling $s$ and $u$ quarks are Cabibbo
suppressed relative to the interactions coupling $d$ and $u$ quarks.
For the nonequilibrium processes such as the conversion from two-flavor
to three-flavor quark matter, however, as the strange quark semi-leptonic 
processes in neutrino emissions and non-leptonic 
processes in axion emissions are of significance. 
 

The axion is a pseudo-Goldstone boson which was introduced by Peccei and
Quinn (PQ) to solve the strong CP-problem in a natural way \cite{PQ}\@.
However, theoretical and experimental investigations 
give little guidance on the PQ symmetry-breaking scale, $f_a$,
and therefore on the mass of the axion \cite{WW}\@.
The axion decay constant $f_a$ is related to the axion mass \cite{axion}  
\[
m_a \simeq \left( \frac{0.62 \times 10^{7} GeV}{f_a} \right) \; eV .
\]
There are two generic types of invisible axions; the
Dine-Fischler-Srednicki-Zhitnitskii (DFSZ) axion which couples to both quarks 
and leptons at tree level \cite{DFSZ} and 
the hadronic, or Kim-Shifman-Vainshtein-Zakharov (KSVZ),
axion which has no tree-level coupling to leptons but does couple to them
at loop levels \cite{KSVZ}\@. 
In invisible axion models, the axion mass $m_a$ is in principle arbitrary, 
however astrophysical and cosmological considerations \cite{axion}
can provide an upper and lower bounds for $m_a$\@.
The astrophysical bounds on $m_a$ are due to the fact that axion emission
is an additional energy loss mechanism for stars.

If such axions could be copiously produced during the conversion of
a neutron star into a strange star, it might drastically alter the 
energy budget of stars. The axions might thus deplete the stellar energy and 
change the usual course of stellar evolution.
The emission of axions would have hastened the cooling process.
The quarks are expected to be highly relativistic and degenerate in such 
matter with their Fermi energies in the range of 300 to 500 $MeV$\@.
A calculation of the axion energy flux from strange quark matter has been 
made by Anand et al.\ \cite{anand}\@. They have shown that the axion emission
rate is 
several orders of magnitude smaller compare to the neutrino emission rate.

We investigate here the energy loss of the neutron star during the conversion 
of non-strange quark matter into strange quark matter in the interior of the 
star. We assume that the axions escape freely 
out of the star as soon as they are produced, and that the temperature in 
the interior is constant during the conversion for the reasons that strong 
and electromagnetic interaction timescales are much smaller than those of weak 
interactions. 
Unless otherwise noted all equations assume $\hbar = c = k_B = 1$\@.


\section{Axion Emission Rate}

The axion is the hypothetical pseudo-Goldstone boson associated with the
spontaneous breakdown of the PQ symmetry proposed to solve the
strong $CP$-problem of QCD\@. We consider the invisible axions proposed by
DFSZ that couple to both quarks and leptons at tree level.
The interaction of axions with quarks can be written as \cite{axion} 
\be
{\cal L}_{int} = \left( \frac{C_i}{2 f_a} \right) 
\bar{q_i} \gamma_{\mu}\gamma_{5} q_{i} \partial^{\mu} a ,
\ee
where $a$ is the pseudoscalar axion field, $f_a$ is the axion decay constant.
The $C_{i}$'s are model-dependent coefficients typically of order one.
At tree level, while $C_{i} = 0$ in the KSVZ axion model, they are
nonzero for the DFSZ axion model. For example \cite{sred},
\ba
C_u &=& -2.76 - cos2\beta , \nn \\
C_d &=& -1.13 - cos2\beta , \\ 
C_s &=& 0.89 + cos2\beta , \nn 
\ea
where $\beta$ depends on the Higgs sector.

Since the conversion of a neutron star into a strange star occurs via
the strangeness-changing non-leptonic weak reaction \cite{madsen}, 
the dominant axion emission process during the conversion period is
\be
u_1 \; + \; d \; \rightarrow \; u_2 \; + \; s \; + \; a \,.
\ee
For this axion emission process, there are four diagrams to be evaluated and 
one of them is shown in Figure 1.
The matrix element from the diagram of Fig.\ 1 can be written as
\be
S_{fi}=\bar{u}_2 \left[ \gm_{\mu} \gm_{5} \left(\frac{C_u}{2 f_a}\right) 
q^{\mu} \; i S_F \; \frac{g}{\sqrt{2}} \gm_{\nu} \frac{1}{2}(1 - \gm_{5}) 
cos\theta_c \right] u_{d} \; D_F \; \bar{u}_s \left[\frac{g}{\sqrt{2}} 
\gm^{\nu} \frac{1}{2}(1 - \gm^{5}) sin\theta_c \right] u_{1} ,
\ee
%
%
\noindent
where the subscripts 1 and 2 in $u$ denote the initial and final $u$ quark
respectively, $g$ is dimensionless weak coupling, $q$ is the axion 
momentum and $\theta_c$ is the Cabbibo angle. The propagators are given as  
\[
iS_F = \frac{i}{\not\!p_2 + \not\!q - m_q}, \;\;\;\;\;
D_F = \frac{1}{k^2 - M_{W}^{2}},
\]
where $p$ is the momentum of the quark and
$k$ is the quark four-momentum transfer $p_2 - p_d$ carried by the weak
boson.
Since $q \ll p_F$, where $p_F$ is the quark Fermi momentum, and with the
quark mass neglected,
the matrix element, Eq. (4), simplifies to 
\be
S_{fi} \simeq i \left(\frac{C_{u}}{2 f_a}\right) \frac{G_F}{\sqrt{2}} 
sin\theta_c cos\theta_c [ \bar{u}_2 \gm_\mu \gm_5 (1-\gm_5) u_d \; 
\bar{u}_s \gm^{\nu}(1-\gm^5) u_1 ] . 
\ee 
Here we used the relation, for $M_{W}^{2} >> k^2$ \cite{HM},
\[
\frac{g^2}{8 M_{W}^{2}} = \frac{G_F}{\sqrt{2}}
\]
where $G_F$ is the Fermi weak coupling constant and $M_W$ is the $W$-boson mass.
For all processes the matrix element squared and summed over the color 
and spins is  
\be
|S_{fi}|_{T}^{2} = 2^3 \left(\frac{G_{F}}{f_a}\right)^2  sin^{2} \theta_c 
cos^{2} \theta_c \{ 2 C_{u}^{2} + C_{d}^{2} + C_{s}^{2} \} |M|^2 ,
\ee
where 
\ba
|M| &=& \bar{u_2} \gm_\mu (1-\gm_5) u_d \; 
\bar{u_s} \gm^{\nu}(1-\gm^5) u_1  \nonumber \\
&=& p_s \cdot p_2 p_d \cdot p_1 \,. \nonumber
\ea

The axion emissivity, the energy loss rate per unit volume and unit time 
due to the axion emission process, can be written as a multi-dimensional 
integral,
\be
{\cal E}_a = \frac{1}{2} \int (6 \times 6) d\Pi_{i} E_a |S_{fi}|_{T}^{2} 
\delta^4 (p_1 + p_d - p_2 - p_s - p_a) {\cal S} ,
\ee
where $i = u_1, d, u_2, s, a$ and 
$d\Pi_{i} = \frac{d^{3}p_i}{(2 \pi)^3 2 E_i}$ is the Lorentz-invariant 
phase-space element. The factor 6 stems from 2 spins and 3 colors, and
$\frac{1}{2}$ is explicitly shown in Eq. (7) to take into account that only
left-handed helicity states of the $u_1$ quark couple to $W$ boson which
mediates the transformation. Henceforth we neglect the masses of 
$u, \; d, \; s$ quarks \cite{madsen}\@.  The statistical factor is
\be
{\cal S} = f_1 f_d (1 - f_2)(1 - f_s).
\ee
Here the functions $f_i$ are the usual Fermi-Dirac distribution functions
of the particle species $i$,
\[
f_i = \frac{1}{e^{\frac{E_i - \mu_i}{T}} + 1} ,
\]
where $\mu_i$ are the chemical potentials usually of order of a few hundred
$MeV$ in the systems of interest. 
The distribution function depends on both the particle momentum and the
species. The combination of distribution functions in Eq. (8) expresses the 
physical fact that for the process to occur,
the initial $u$ and $d$ quark states must be occupied and the final
$u$ and $s$ quark states unoccupied \cite{iwamoto}\@.
In quark matter, strong and electromagnetic interaction processes establish
thermal equilibrium at a rate much faster than that of weak interaction
processes. Therefore, we take the equilibrium Fermi-Dirac distribution 
function.

In order to evaluate the energy loss rate Eq. (7),
we first note that $|\vec{q}|$ is small compared to the $\vec{p}_i$'s 
and is therefore ignored in momentum conservation. Since highly degenerate 
relativistic quark matter enables us to replace the momenta by their values at
the top of the Fermi sea, the quark momenta 
$|\vec{p}_i|$'s are set equal to their respective Fermi momenta $p_{F}(i)$\@. 
We shell neglect the effects of interactions on the 
particle density of states. Then, the phase space element 
$d^3p_i = |\vec{p}_i|^2 d|\vec{p}_i| d\Omega_i$, where $d\Omega_i$ is an 
element of solid angle, becomes $d^3p_i = |\vec{p}_i|^2 dE_i d\Omega_i$\@.

In an integral over energies \cite{burrows}, we obtain
\ba
\int \Pi_{i} d E_i \delta (E_1 + E_d - E_2 - E_s - E_a) {\cal S} \nn
= \frac{Q}{6} \; \frac{Q^2 + (2 \pi)^2 T^2}{e^{\frac{Q}{T}} - 1}
\ea
where
\[
Q = E_a + \delta \mu, \;\;\;\;\;\; \delta \mu = \mu_d - \mu_s .
\]
We now integrate the above quantity over axion energies;
\ba
{\cal T}(T, \delta\mu)&\equiv&\int dE_a E_{a}^{2} Q \frac{Q^{2}+(2 \pi)^2 
T^2}{e^{\frac{Q}{T}}-1} \nn \\
&=& \frac{62}{945} T^6 
- \frac{8}{3 \pi^6} T^5 \delta \mu [3 \zeta(5) + \pi^2 \zeta(3)]
+ \frac{11}{90 \pi^2} T^4 \delta \mu^2 \,.\nn 
\ea
For an integral over angles \cite{shapiro},
\ba
{\cal A} &\equiv& \int \Pi_{j} d\Omega_j  \delta^3
(\vec{p_1} + \vec{p_d} - \vec{p_2} - \vec{p_s}) |M|^2 \nn \\
&=& \frac{1}{p_{F}^{3}} \frac{(4 \pi)^5}{2} \,.
\ea
where we ignored the axion momentum, and $j = u_{1}, d, u_{2}, s$\@.
Therefore, Eq. (7) is written as 
\be
{\cal E}_a =\frac{2^4 (2\pi)^4}{[2 (2 \pi)^3]^5}\left(\frac{G_F}{f_a}\right)^2 
\{2 C_{u}^{2} + C_{d}^{2} + C_{s}^{2}\}
sin^{2}\theta_c cos^{2}\theta_c p_{F}^{8} \; {\cal A}\;{\cal T}(T, \delta\mu) .
\ee
Finally, from Eqs. (9) and (10), we obtain the axion emissivity
\ba
{\cal E}_a &=& 1.5 \times 10^{32} \{2 C_{u}^{2} + C_{d}^{2} + C_{s}^{2} \} 
sin^{2} \theta_c cos^{2} \theta_c \left(\frac{GeV}{f_a}\right)^2 
\left(\frac{p_{F}}{500 MeV}\right)^{5} {\cal T}_{MeV}(T, \delta\mu) \nn \\
&~& ergs/cm^{3}/sec \,.
\ea
where
\ba
{\cal T}_{MeV}(T, \delta\mu) &=& \frac{62}{945} \left(\frac{T}{MeV}\right)^6
-\frac{8}{3 \pi^6} \left(\frac{T}{MeV}\right)^5 \left(\frac{\delta \mu}{MeV}
\right) [3 \zeta(5) + \pi^2 \zeta(3)] \nn \\ 
&~& + \frac{11}{90 \pi^2} \left(\frac{T}{MeV}\right)^4 
\left(\frac{\delta \mu}{MeV}\right)^2 .  
\ea
Figure 2 shows ${\cal T}_{MeV}(\delta\mu)$ as a function of $\delta\mu$
for the fixed temperatures $T=5,\; 10$ and $20 MeV$\@. 
For the chemical equilibrium, $\delta \mu = 0$, then 
${\cal T}_{MeV} \simeq 6 \times 10^4$ for $T=10 MeV$\@. However,
for the typical value of $\delta\mu =300 MeV$, ${\cal T}_{MeV} \simeq  10^7$ 
for $T=10 MeV$\@.

%





\section{Results and Discussions}

Our main result for the total energy flux rate for the emission of axions
from conversion of non-strange quark matter into strange quark matter is given
by Eq. (12). 
Here we can carefully quote our bound on the axion coupling $f_a$, 
using the simple inequality \cite{ishizuka}, 
\be
{\cal E}_a \cdot V \cdot \delta t < E_m
\ee
The allowed energy loss, $E_m / \delta t$ is taken to be 
$10^{52}$ erg/sec for definiteness, and the volume of axion emitting
region $V = \frac{4}{3} \pi R^3$ with $R = 10 \sim 20 km$\@. 
As for the conversion timescale, if $T=10 MeV$, it varies roughly
between one and ten minutes \cite{olinto}\@. So, we take the timescale
typically $\delta t = 100 sec$\@. Then, from Eq. (14), we can obtain the 
bound on $f_a$,
\be
\sim 10^5 \; GeV < f_a \,.
\ee
We note that the bound based on Eq. (14) is legitimate only if the emitted
axions never interact on their way out of the star. If the axion coupling is 
large enough, axions once produced may interact many times, namely be absorbed
and reemitted more than once in the hot core of the star, and may be
trapped thermally. 

As applications, Mikheev and Vassilevskaya \cite{mik} recently investigated 
the radiative decay of the axion $a \rightarrow \gamma\gamma$ in an external 
electromagnetic field in the DFSZ model.
They concluded that the decay probability is strongly catalyzed by the external
field, namely, the field removes the main suppression caused by the smallness
of the axion mass. Therefore, the radiative decay of the axion in strong
magnetic fields of the neutron star could give interesting astrophysical 
phenomena.
Furthermore, we can consider that the sudden conversion from a neutron star to 
a strange star may account for the gamma ray bursts at comological distances 
\cite{olinto}\@. 
Most of the energy is probably released in the form of neutrinos.
If a part of the total energy goes into $\gamma$ rays decayed from the axions, 
it will be large enough
to account for the gamma ray bursts at cosmological distances and to explain 
their isotropic distribution. 
The outcome of such a conversion event will be the emission of as much as
$\sim 10^{58} MeV$ of energy in $\sim sec$ to $\sim min$ as radiation with a
typical temperature of tens of $MeV$\@. This conversion can be observed as 
a gamma
ray burst \cite{olinto}\@. 

In summary, as a possible mechanism for production of axions, 
we considered the conversion of non-strange quark matter into strange 
quark matter. 
We also estimated the energy loss of neutron stars through the emission
of axions in addition to the cooling provided by the neutrino emission. 
During the conversion period, the important process is the non-leptonic
weak-interaction.
We assume that the axions are not trapped and escape freely 
out of the star as soon as they are produced, and that the temperature in 
the interior is constant during the conversion.
We find that the energy carried away by axions during the conversion 
is not negligible. It is also found that the axion emission rate is three orders
of magnitude larger compared to the chemical equilibrium case.
In addition, we discussed the bound on the axion coupling.

\vspace*{3cm}

\acknowledgments

\noindent
We would like to thank Prof. N. Iwamoto and Prof. K. Choi
for their encouraging comments. This work is supported in part by the Basic
Science Institute Program, Korean Ministry of Education (Project 
No. BSRI-97-2441) and in part by the KOSEF through the Center for Theoretical
Physics of Seoul National University. 


\newpage
\begin{center}
{\bf Figure Captions}
\end{center}

\noindent
Fig. 1 Feynman diagram for axion emission in the non-leptonic weak interaction.
\\

\noindent
Fig. 2 ${\cal T}_{MeV}(\delta\mu)$ as a function of $\delta\mu$
for the fixed temperature $T=10$ MeV. The solid, dash, and dot lines
correspond to $T = 5, \; 10,$ and $20$ MeV, respectively.

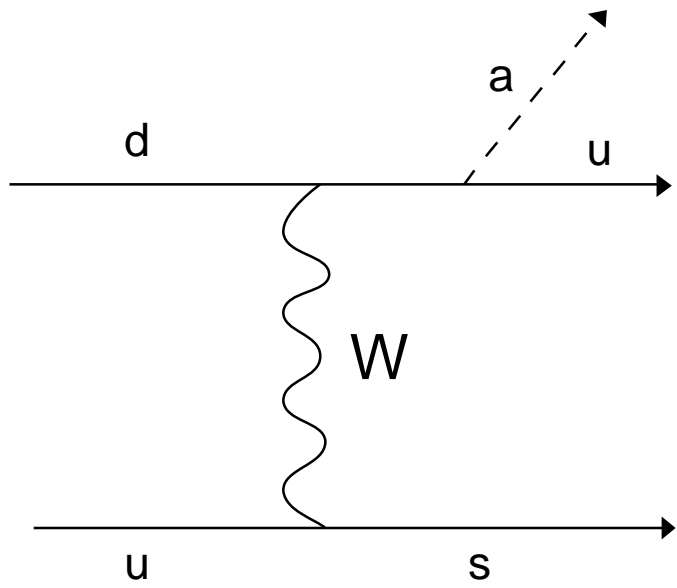
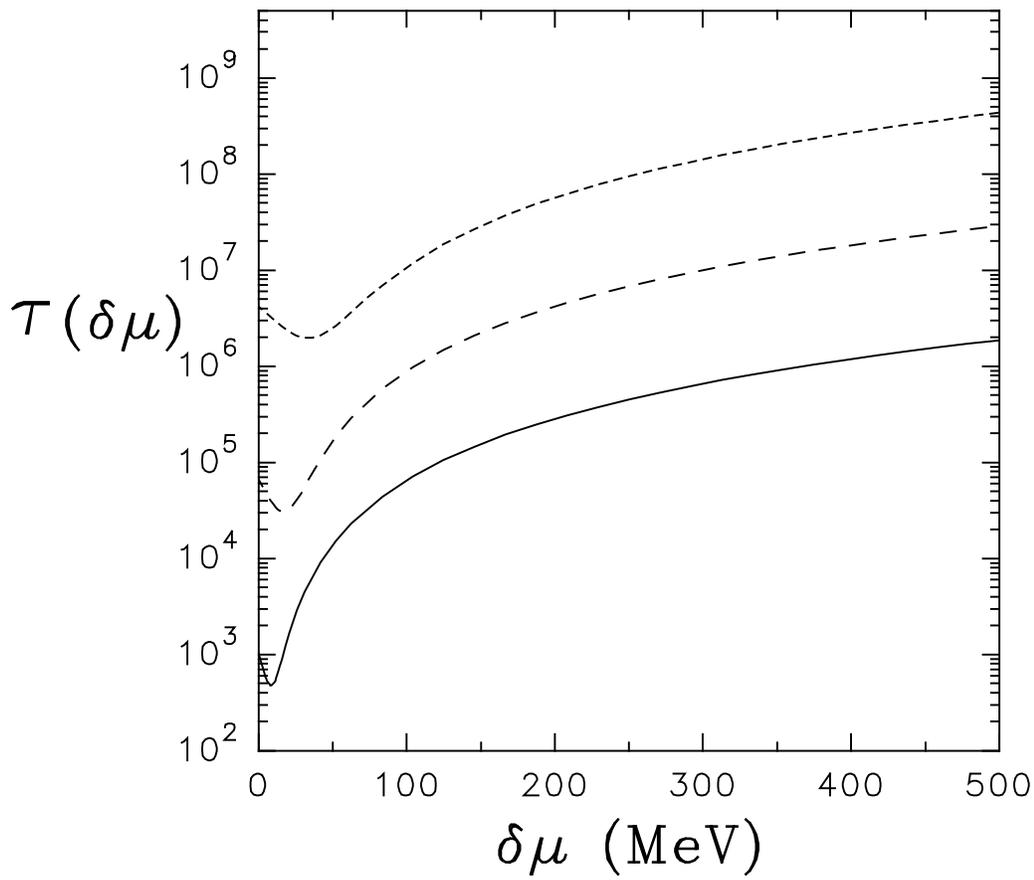
\end{document}